\documentclass{amsart}
\usepackage{amsmath, enumerate}
\usepackage{algorithm}
\usepackage{algpseudocode}
\usepackage{amsfonts}
\usepackage{amssymb}
\usepackage{amsthm}
\usepackage{color}
\usepackage{caption}
\usepackage{graphicx}
\usepackage{url}
\usepackage{verbatim}

\newcommand{\Z}{\mathbb{Z}}

\newcommand{\N}{\mathbb{N}}

\newcommand{\Q}{\mathbb{Q}}

\newcommand{\bp}{\begin{problem}}

\newcommand{\ep}{\end{problem}}

\newcommand{\ba}{\begin{answer}}

\newcommand{\ea}{\end{answer}}

\newcommand{\ben}{\renewcommand{\theenumi}{\alph{enumi}}

\renewcommand{\labelenumi}{(\theenumi)}\begin{enumerate}}

\newcommand{\een}{\end{enumerate}}

\newcommand{\Aut}{\mathrm{Aut}}
\newcommand{\Inn}{\mathrm{Inn}}
\newcommand{\Out}{\mathrm{Out}}
\newcommand{\End}{\mathrm{End}}

\theoremstyle{definition}
\newtheorem{defin}{Definition}[section]
\theoremstyle{plain}
\newtheorem{thm}[defin]{Theorem}

\newtheorem{lem}[defin]{Lemma}
\newtheorem{prop}[defin]{Proposition}

\DeclareCaptionType{result}[Table][List of Results]

\title[Polycyclic Group-Based Cryptography]{The Status of Polycyclic Group-Based Cryptography: A Survey and Open Problems}
\begin{document}

\author[J. Gryak]{Jonathan Gryak}
\address{Jonathan Gryak, CUNY Graduate Center, PhD Program in Computer Science, City University of New York}
\email{jgryak@gradcenter.cuny.edu}

\author[D. Kahrobaei]{Delaram Kahrobaei}
\address{Delaram Kahrobaei, CUNY Graduate Center, PhD Program in Computer Science and NYCCT, Mathematics Department, City University of New York}
\email{dkahrobaei@gc.cuny.edu}

\begin{abstract} Polycyclic groups are natural generalizations of cyclic groups but with more complicated algorithmic properties. They are finitely presented and the word, conjugacy, and isomorphism decision problems are all solvable in these groups. Moreover, the non-virtually nilpotent ones exhibit an exponential growth rate. These properties make them suitable for use in group-based cryptography, which was proposed in 2004 by Eick and Kahrobaei \cite{EickKahrobaei04}.

\

Since then, many cryptosystems have been created that employ polycyclic groups. These include key exchanges such as non-commutative ElGamal, authentication schemes based on the twisted conjugacy problem, and secret sharing via the word problem. In response, heuristic and deterministic methods of cryptanalysis have been developed, including the length-based and linear decomposition attacks. Despite these efforts, there are classes of infinite polycyclic groups that remain suitable for cryptography.

\

The analysis of algorithms for search and decision problems in polycyclic groups has also been developed. In addition to results for the aforementioned problems we present those concerning polycyclic representations, group morphisms, and orbit decidability. Though much progress has been made, many algorithmic and complexity problems remain unsolved; we conclude with a number of them. Of particular interest is to show that cryptosystems using infinite polycyclic groups are resistant to cryptanalysis on a quantum computer.
\end{abstract}

\maketitle
\tableofcontents

\section{Introduction}
In cryptography, many of the most common key exchange protocols, including RSA and Diffie-Hellman, rely upon hardness assumptions related to integer factorization and discrete logarithms for their security. While there are no known efficient algorithms for performing the above operations on conventional computers, Peter Shor devised a quantum algorithm \cite{shor1994} that solves both of these problems in polynomial time. This has motivated the search for alternative methods for constructing cryptosystems. One such methodology is non-commutative cryptography, which unlike the aforementioned conventional systems does not operate over the integers. Instead, non-commutative cryptographic systems are built upon groups and other algebraic structures whose underlying operations are non-commutative.\\
\paragraph{}
In 1999, Anshel, Anshel, and Goldfeld \cite{AAG99} and Ko, Lee, et al. \cite{KoLee00} introduced key exchange protocols whose security is based in part on the conjugacy search problem: for a group $G$, given that $ u,v\in G$ are conjugate, find an $x$ in $G$ such that $u^x=v$.  Though braid groups were the suggested platform for both protocols, other classes of groups can be employed. In general, groups suitable for use in non-commutative cryptography must be well-known and possess the following properties:  a solvable word problem, a computationally difficult group-theoretic problem, a ``fast" word growth rate, and the namesake non-commutativity \cite{Myasnikov-book}.\\
\paragraph{}
In 2004, Eick and Kahrobaei \cite{EickKahrobaei04} investigated the algorithmic properties of polycyclic groups. In particular, they explored how the time complexity of the word and conjugacy problems varied with respect to a group's Hirsch length. Their experiments showed that the while the time complexity of the conjugacy problem grew exponentially with increased Hirsch length, the word problem remained efficiently solvable. These results suggested the suitability of polycyclic groups for use in cryptography, and stimulated research into cryptosystems based on these groups and their underlying algorithmic problems.\\
\paragraph{}
In this paper, we survey the development of group-based cryptography over polycyclic and metabelian groups. In section 2 we discuss the algorithmic properties of polycyclic groups. Polycyclic groups and their intrinsic presentations are defined, as well as several other representations. A number of group-theoretic decision problems are introduced, including the word, conjugacy, and isomorphism decision problems. Note that in every polycyclic group, the three aforementioned problems are solvable. Moreover, the word problem can be solved efficiently in most cases by using a collection algorithm.\\
\paragraph{}
In section 3 we describe a number of cryptosystems that have been built around these groups. These include additional key exchanges along with schemes for secret sharing, authentication, and digital signatures.  This variety of cryptosystems evinces the flexibility and utility of polycyclic groups in non-commutative cryptography.\\
\paragraph{}
As new cryptosystems are created, so too are their dual in the form of cryptanalyses and attacks. In section 4 we discuss the length-based attack, a heuristic technique that was the first to break the AAG protocol over braid groups. Other attacks exploit the linear representation that all polycyclic groups admit. Some, such as the field-based attack, are specific to a subclass of polycyclic groups. A more general approach is the linear decomposition attack, but its feasibility is dependent upon the size of a group's representation.\\
\paragraph{}
We conclude the paper with the current status of polycyclic groups cryptography. We also include a list of open problems, which we hope will guide researchers who wish to work in this exciting field.

\section{Algorithmic Problems in Polycyclic Groups}
\label{algpc}
The nature of polycyclic groups enables them to be represented in several ways. These approaches give rise to complementary algorithms for solving search and decisions problems, with varying degrees of computational complexity. Due to this flexibility, we begin our study of the algorithmic problems in polycyclic groups by examining these representations.

\subsection{Representations of Polycyclic Groups}
\subsubsection{Polycyclic Sequences and Hirsch Length}
A group $G$ is said to be {\em polycyclic} if it has a subnormal series $G=G_1 \triangleright \cdots \triangleright G_{n+1}=\{1\}$ such that the quotient groups $G_i/G_{i+1}$ are cyclic. This series is called a \emph{polycyclic series}. The {\em Hirsch length} of a
polycyclic group $G$ is the number of infinite groups in its
polycyclic series. Though a polycyclic group can have more than one polycyclic series, as a consequence of the Schreier Refinement Theorem, its Hirsch length is independent of the choice of series.

\subsubsection{Polycyclic Presentations}
Every polycyclic group can be described by a polycyclic presentation:
$$
\begin {array}{lrl}
\langle g_1,\ldots,g_n \mid &
g_j^{g_i}=u_{ij}		& \text{for} \; 1\leq i<j\leq n,\\
&g_j^{g_i^{-1}}=v_{ij}	& \text{for} \; 1\leq i<j\leq n,\\
&g_i^{r_i} =w_{ii}	& \text{for} \; i \in I \rangle,
\end{array}
$$
where $u_{ij},v_{ij},w_{ii}$ are words in the generators $g_{i+1},\ldots ,g_n$ and $I$ is the set of indices $i \in \{1,\ldots,n\}$ such that $r_i=[G_i:G_{i+1}]$ is finite.

\

This special type of finite presentation reveals the polycyclic structure of the underlying group, see
\cite[Chapter 10]{HEO05} for details. Unlike general finite 
presentations, a polycyclic presentation enables the word problem to be solved
using an algorithm called {\em collection}. The collection algorithm is generally effective in practical applications, but its precise computational complexity remains unknown. For finite groups, 
collection from the left was shown to be polynomial by Leedham-Green and Soicher \cite{LGS90}. For infinite groups, the complexity of the collection algorithm and a modified version were analyzed by Gebhardt 
\cite{Geb02}. The resultant worst-case bound is in terms of the absolute values of all 
exponents occurring during the collection process, rather than the exponents of the input word.
Thus a global complexity analysis of the collection algorithm remains elusive.

\subsubsection{Polycyclic Presentations with a Malcev Basis} It has been shown
by Assmann and Linton \cite{ALi07} that the efficacy of the 
collection algorithm can be improved significantly by exploiting the
Malcev structure of the underlying group. This approach determines
a large nilpotent normal subgroup of the given group and then exploits
the Malcev correspondence for the normal subgroup. There is no known
complexity analysis for this methodology.

\subsubsection{Polycyclic Presentations with Multiplication Polynomials} Du Sautoy 
\cite{duS02} proved that every polycyclic group has a normal subgroup
of finite index such that multiplication in this subgroup can
be achieved by evaluating certain multiplication polynomials. This extends
the well-known result by Hall \cite{Hal69} for torsion-free nilpotent
polycyclic groups. If such multiplication polynomials are available
the performance of collection in the considered group improves significantly. Additionally, 
it provides a basis for the complexity analysis of
multiplication in polycyclic groups; it must be noted however that
the index of the normal subgroup can be arbitrarily large.

\subsubsection{Matrix Groups}
\label{matrixrep}
It is well-known that every polycyclic group can 
be embedded into $GL(n, \Z)$ for some $n \in \N$. For groups that are additionally torsion-free and nilpotent, a matrix representation can be computed. The algorithm of
Lo and Ostheimer \cite{LOs99} can be applied to a polycyclic presentation, while for multiplication polynomials the technique by Nickel \cite{Nic06} can be utilized. Multiplication of group elements in their matrix form is polynomial in the dimension $n$ of the representation.

\subsection{Growth Rate}\label{growthrate}
\paragraph{}
Let $G$ be a finitely generated group.  The growth rate of a group is specified by its \emph{growth function}
$\gamma: \mathbb{N} \longrightarrow \mathbb{R}$  defined as
$\gamma(n) = \# \{ w \in G : l(w) \leq n \}$, where $l(w)$ is the
length of $w$ as a word in the generators of $G$. As words are used as keys in group-based cryptography, there is a natural relationship between the growth rate of a group and the \emph{key space}, the set of all possible keys. A fast growth rate engenders a large key space, making an exhaustive search
of this space intractable.

\

A large class of polycyclic groups are known to have an exponential growth rate (namely those which are not virtually nilpotent, see Wolf \cite{Wolf} and Milnor \cite{Milnor}). Consequently, these polycyclic groups are potentially good candidates for use as platform groups.

\subsection{Decision Problems}
\paragraph{}
In 1911, Max Dehn introduced \cite{dehn1911} three decision problems on finitely presented groups - the word problem, the conjugacy problem, and the isomorphism problem. In the definitions below, let $G$ be a finitely presented group:
\begin{itemize}
\item \emph{Word Decision Problem} - For any $g\in G$, determine if $g=1_G$, the identity element of $G$.
\item \emph{Single Conjugacy Decision Problem} - Determine for any $u,v\in G$ if $u$ is conjugate to $v$ (denoted $u\sim v$).
\item \emph{Isomorphism Decision Problem} - Given groups $G$ and $G'$ with respective finite presentations $\langle X\mid R\rangle$ and $\langle X'\mid R'\rangle$, determine if $G$ is isomorphic to $G'$.
\end{itemize}

For polycyclic groups all three of the above problems are decidable. The conjugacy decision problem for polycyclic groups is decidable by the results of Remeslennikov \cite{remeslennikov1969conjugacy} and Formanek \cite{formanek1976conjugate}. That the word problem is decidable can be observed from its formulation as a special case of the conjugacy decision problem (where $g=u,v=1_G$), or by observing that every word has a unique normal form induced by a polycyclic presentation. The isomorphism decision problem for polycyclic groups is solvable by a result of Segal \cite{Segal90decidable}.

\

An additional decision problem called the \emph{subgroup membership decision problem} (alternatively the generalized word decision problem) asks for any $g\in G$ and subgroup $H\leq G$, determine if $g\in H$. Malcev in \cite{malcev1983homomorphisms} showed that this problem is indeed solvable for polycyclic groups.

\subsection{The Conjugacy Search Problem and its Variations}
\paragraph{}
Once the solvability of a group-theoretic decision problem is affirmed, the subsequent task is to produce elements (or morphisms, etc.) that are solutions to particular instances of it. The seminal protocols of non-commutative cryptography, Ko-Lee and AAG, are based in part on the conjugacy search problem (CSP). Their example spurred the development of many other protocols whose security is based on some variant of the CSP.
In this section we explore these variations and the methods designed to solve them.

\subsubsection{Conjugacy Search Problem}
Let $G$ be a group and $a_1, \ldots, a_n, b_1, \ldots, b_n$ elements of it, with $a_i\sim b_i$. The problem of finding a $c\in G$ such that for all $i$, $a_i^c=b_i$ is called the \emph{(single) conjugacy search problem} for $i=1$ and the \emph{multiple conjugacy search problem} for $1< i \leq n$. In polycyclic groups, the multiple conjugacy search problem for $n$ elements reduces to $n$ independent solutions of single conjugacy search \cite{EickKahrobaei04}. We will therefore speak only of the conjugacy search problem without signifying arity.

\

For any finitely presented group (polycyclic groups included) the conjugacy search problem can be solved exhaustively by recursively enumerating the conjugates of the element in question \cite{shpilrain2010search}.
There are other approaches to solving the conjugacy search problem, many of which can solve it efficiently. However, the applicability of these methods and their relative efficiency is contingent upon addition restrictions on the group's properties, as well as the manner is which the polycyclic group is specified.

\subsubsection{CSP Using Polycyclic Presentations}

\label{orbstab}
For infinite polycyclic groups the algorithm proposed by Eick and Ostheimer \cite{EOs03} is applicable. This algorithm uses a variety of ideas: it exploits finite orbit and stabilizer computations, calculations in number fields, and linear methods for polycyclic groups. The algorithm has been implemented and seems to be efficient for groups of small Hirsch length. An analysis of the algorithm's complexity is hindered by there being no bound on the length of the finite orbits that may occur in the computation.

\

The restriction of the applicability of the above algorithm to groups of small Hirsch length is supported by the experimental evidence provided by Eick and Kahrobaei in \cite{EickKahrobaei04}. They compared the performance of the Eick-Ostheimer algorithm for the CSP against the collection algorithm for polycyclic groups of the form $G=\mathcal{O}_K\rtimes \mathcal{U}_K$, where $\mathcal{O}_K$ and $\mathcal{U}_K$ are respectively the maximal order and group of units of an algebraic number field $K$. In the table below, the column $H(G)$ is the Hirsch length of the group $G$, with the collection and conjugation entries representing the average running time over 100 trials using random words (respectively, random conjugate pairs) from $G$:
\begin{table}[htb]
\begin{center}
\begin{tabular}{ccr}
H(G) &  Collection & Conjugation\\
\hline
2   & 0.00 sec &  9.96 sec \\
6   & 0.01 sec & 10.16 sec \\
14 & 0.05 sec & $>\!100$ hr\;
\end{tabular}
\end{center}
\end{table}

These results suggest that while collection remains efficient as the Hirsch length increases, the Eick-Ostheimer algorithm becomes impractical. Presently there are no known algorithms for infinite polycyclic groups of high Hirsch length. Such groups remain suitable for use as platform groups.

\subsubsection{CSP Using Multiplication Polynomials}

Suppose that $G$ instead is given by a polycyclic presentation with multiplication
polynomials. Let $g_1, \ldots, g_k$ be the polycyclic generating set 
of the presentation and consider a generic element $g = g_1^{x_1} \cdots
g_k^{x_k}$ of $G$. $g$ is a solution to the multiple conjugacy search problem
if and only if $a_i g = g b_i$ for $1 \leq i \leq k$. 
If $a_i = g_1^{l_{i1}} \cdots g_k^{l_{ik}}$ and  $b_i = 
g_1^{m_{i1}} \cdots g_k^{m_{ik}}$, with $f_1, \ldots, f_k$ denoting the
multiplication polynomials for $G$, then $a_i g = g b_i$ if and only if
\[ f_j(l_i, x) = f_j(m_i,x) \mbox{ for } 1 \leq i,j \leq k.\]
If $f_1, \ldots, f_k$ are given as explicit polynomials over an extension
field of $\Q$ and $l_i, m_i$ are integer vectors, then the CSP is equivalent to
determining an integer solution for a set of $k$ polynomials in $k$ 
indeterminates. Thus the CSP can also be considered from the perspective of 
algebraic geometry.

\subsubsection{Power Conjugacy Search Problem}
The key exchange presented in Section \ref{kkpower} makes use of the \emph{power conjugacy search problem}, where if it is known for some $a,b\in G$  and $n \in \N$  that $a^n = b^g$ for some $g \in G$, the task is to find one such $n$ and $g$. Note that for $n=1$ this reduces to the standard CSP, whereas if $g=1_G$ this reduces to the \emph{power search problem}.

\

Just as the conjugacy search problem is solvable by enumeration, so is the power conjugacy search variant, but no efficient algorithm is known.

\subsubsection{Twisted Conjugacy Search Problem}
Twisted conjugacy arises in Nielsen theory, where the number of twisted conjugacy classes is related to number of fixed points of a mapping. The \emph{twisted conjugacy search problems} is to find, given a group $G$ and an endomorphism $\phi$, an element $a \in G$ such that $t = a^{-1}w \phi(a)$, provided that at
least one such $a$ exists. 

\

The standard CSP can be seen as a special case of the twisted version where $\phi(x)=x$, the identity automorphism. The protocol by Shpilrain and Ushakov in Section \ref{twistedauth} uses the double twisted conjugacy variant, in which the above definitions is modified to include an additional endomorphism $\alpha$ and the task is to then find an element $a \in G$ such that $t = \alpha(a^{-1})w \phi(a)$.

\

The twisted conjugacy decision problem was proven to be decidable by Roman'kov \cite{roman2010twisted}. Both the single and doubly twisted conjugacy search problems are solvable by the same method of enumeration as in the case of the standard conjugacy search problem. However, no efficient algorithm is known.

\subsection{Properties of Automorphism Groups}
 The automorphism group $\Aut(G)$ and its subgroups have been extensively studied for polycyclic groups $G$. Like polycyclic groups themselves, $\Aut(G)$ is finitely presented \cite{auslander1969automorphism}, and the outer automorphism group $\Out(G)$ is isomorphic to a linear group \cite{wehrfritz1994two}.

\

 A decision problem related to $\Aut(G)$ is the \emph{orbit decision problem}. Given elements $g,h\in G$ and a subset $A\subseteq\Aut(G)$, determine if there exists $\alpha\in A$ such that $g=\alpha(h)$. Note that if $A=\Inn(G)$ this problem reduces to the standard conjugacy decision problem.  When $G$ is polycyclic all cyclic subgroups  $A\leq\Aut(G)$ are orbit decidable \cite{bogopolski2010orbit}.

\

For groups $G$ in the larger class of polycyclic-by-finite (or virtually polycyclic) groups, the conjugacy decision problem is decidable in $\Aut(G)$ \cite{Segal90decidable}. Additionally, $\Aut(G)$ is either virtually polycyclic or it contains a non-abelian free subgroup \cite{eick2003automorphism}.

\subsection{Quantum Algorithms}
\paragraph{}
As mentioned in the introduction, the introduction non-commutative cryptography was spurred by the publication of Shor's algorithm. The algorithm enables a sufficiently sized quantum computer to perform integer factorization and compute discrete logs in polynomial time, as opposed to in exponential time on a conventional computer. 

\

From a group-theoretic perspective, Shor's algorithm can be seen as solving the \emph{hidden subgroup problem} in finite cyclic groups. A subgroup $H\leq G$ is considered \emph{hidden} by a function $f$ from $G$ to a set $X$ if  it constant over all cosets of $H$. A 2003 paper by \cite{batty2003quantum} by Batty, et al. explores this and other applications of quantum algorithms to group theory, including an algorithm by Watrous that determines the order of a finite solvable group. Bonanome showed \cite{bonanome2007} that a modified version of Grover's algorithm can solve the automorphism and conjugacy decision problems in finite groups, as well as determine fixed points. The algorithm by Ivanyos, et al \cite{ivanyos2008efficient} solves the hidden subgroup problem for finite nilpotent groups of class 2. There are also partial results to solving the power conjugacy problem \cite{fesenko2014vulnerability}.

\

Despite these developments in the use quantum algorithms for finite groups, there are no known quantum algorithms that are applicable to infinite groups.

\section{Cryptosystems}
For the systems described below, the chosen platform group $G$ should be suitable for cryptography as delineated in the introduction. Let $G$ be finitely presented and non-abelian. Group operations (products, inverses) and solving the word problem must be efficient. Additional criteria for each protocol or scheme are stated in their respective descriptions. Note that the precise definitions of each algorithmic search or decision problem can be found in Section \ref{algpc}.
\subsection{The Anshel-Anshel-Goldfeld Key-Exchange Protocol}
\label{AAG}
In their 1999 paper \cite{AAG99}, Anshel, Anshel, and Goldfeld introduced the \emph{commutator key exchange protocol}, which is also referred to as AAG key exchange or Arithmetica. The group-based version of the key exchange described below is in the style of \cite{MyasnikovUshakov07}. Prior to the key exchange, the protocol parameters $N_1,N_2,L_1,L_2,L\in\N$, with $1\leq L_1\leq L_2$, are chosen and made public:
\begin{enumerate}
\item Alice chooses a set $\bar{A}=\{a_1,\ldots,a_{N_1}\}$, with Bob choosing $\bar{B}=\{b_1,\ldots,b_{N_2}\}$, where $a_i,b_j\in G$ are words of length in $[L_1, L_2]$. Note that $\bar{A}$ and $\bar{B}$ both generate subgroups of $G$. These sets are then exchanged publicly with each other.
\item Alice constructs her private key as $A=a_{s_1}^{\epsilon_1}\ldots a_{s_L}^{\epsilon_L}$ , with $a_{s_k}\in \bar{A}$ and $\epsilon_k\in\{-1,1\}$. Similarly, Bob computes as his private key $B=b_{t_1}^{\delta_1}\ldots b_{t_L}^{\delta_L}$, with $b_{t_k}\in \bar{B}$ and $\delta_k\in\{-1,1\}$.
\item Alice then computes $b_j'=A^{-1}b_jA$ for $1\leq j\leq N_2$ and sends this collection to Bob, while Bob computes and sends Alice $a_i'=B^{-1}a_iB$ for $1\leq i \leq N_1$.
\item Alice and Bob can now compute a shared key $\kappa=A^{-1}B^{-1}AB$, which is the \emph{commutator} of $A$ and $B$, denoted $[A,B]$. Alice computes (using only the $a_i'$ which correspond to some $s_i$ of her private key):
\begin{align*}
\kappa_A&=A^{-1}{a'}_{s_1}^{\epsilon_1}\cdots {a'}_{s_L}^{\epsilon_L}\\
	&=A^{-1}B^{-1}a_{s_1}^{\epsilon_1}B\cdots B^{-1}a_{s_L}^{\epsilon_L}B\\
	&=A^{-1}B^{-1}a_{s_1}^{\epsilon_1}(BB^{-1})a_{s_2}^{\epsilon_2}B\cdots B^{-1}a_{s_{L-1}}^{\epsilon_{L-1}}(BB^{-1})a_{s_L}^{\epsilon_L}B\\
	&=A^{-1}B^{-1}a_{s_1}^{\epsilon_1}a_{s_2}^{\epsilon_2}\cdots a_{s_{L-1}}^{\epsilon_{L-1}}a_{s_L}
^{\epsilon_L}B\\
	&=A^{-1}B^{-1}AB.
\end{align*}
Analogously, Bob computes $\kappa_B=B^{-1}A^{-1}BA$. The shared secret is then $\kappa=\kappa_A=\kappa_B^{-1}$.
\end{enumerate}
As noted in \cite{shpilrain2006conjugacy}, the security of AAG is based on both the simultaneous conjugacy search problem and the subgroup membership search problem.
\subsection{Ko-Lee Key Exchange Protocol}
Originally specified by Ko, Lee, et al. \cite{KoLee00} using braid groups, their non-commutative analogue of Diffie-Hellman key exchange can be generalized to work over other platform groups. Let $G$ be a finitely presented group, with $A,B \leq G$ such that all elements of $A$ and $B$ commute. 
\

An element $g\in G$ is chosen, and $g, G, A, B$ are made public. A shared secret can then be constructed as follows:
\begin{itemize}
\item Alice chooses a random element $a\in A$ and sends $g^{a}$ to Bob.
\item Bob chooses a random element $b\in B$ and sends $g^{b}$ to Alice.
\item The shared key is then $g^{ab}$, as Alice computes $(g^b)^a$, which is equal to Bob's computation of $(g^a)^b$ as $a$ and $b$ commute.
\end{itemize}

The security of Ko-Lee rests upon solving the conjugacy search problem within the subgroups $A, B$.

\subsection{Non-Commutative ElGamal Key-Exchange}
In the 2006 paper by Kahrobaei and Khan \cite{KK06}, the authors proposed two adaptations of the ElGamal asymmetric key encryption algorithm for use in non-commutative groups. Let $S,T$ be finitely generated subgroups such that all elements of $S$ and $T$ commute. In any exchange, the triple $\langle G, S, T\rangle$ is made public.

\subsubsection{Non-Commutative Key Exchange Using Conjugacy Search}

\begin{itemize}
\item Bob chooses $s\in S$ as his private key, a random element $b\in G$, and publishes as his public key the tuple $\langle b, c\rangle$, with $c=b^s$.
\item To create a shared secret $x\in G$, Alice chooses $x$ and a $t\in T$. Using Bob's public key, she publishes $\langle h, E\rangle$, with $h=b^t$ and $E=x^{c^t}$.
\item To recover $x$, Bob first computes $h^s$, which, as elements of $S$ and $T$ commute, yields
$$
h^s=(b^t)^s=(b^s)^t=c^t.
$$
Bob can then calculate $x=E^{(c^t)^{-1}}$.
\end{itemize}

The security of this scheme relies upon the conjugacy search problem in $G$.

\subsubsection{Non-Commutative Key Exchange Using Power Conjugacy Search}\label{kkpower}
By imposing the additional requirement that the conjugacy search problem is efficiently solvable in $G$, we can now describe a variation of the previous protocol:
\begin{itemize}
\item Bob chooses $s\in S$ and $n\in\Z$ as his private key, as well as a random element $b\in G$, and publishes as his public key $\langle v,w\rangle$, with $v=g^n$ and $w=g^{-1}sg$. Note that $w^n=(s^{-1}gs)^n=s^{-1}g^ns=s^{-1}vs$.
\item Alice chooses a shared secret $x \in G$, along with $m\in\Z$ and $t\in T$, and publishes $\langle h, E\rangle$, with $h=t^{-1}w^mt$ and $E=x^{-1}t^{-1}v^mtx$.
\item To recover $x$, Bob first computes $E'=sh^ns^{-1}=st^{-1}sg^{mn}st$, which, as elements of $S$ and $T$ commute, yields
$$
E'=t^{-1}v^mt.
$$
Knowing that $E=x^{-1}E'x$, Bob can then solve the conjugacy search problem to obtain the shared secret $x$.
\end{itemize}

The security of this scheme rests upon the power conjugacy search problem in $G$.

\subsection{Non-Commutative Digital Signature}
The following digital signature scheme was proposed in a paper by Kahrobaei and Koupparis \cite{KK12}. The platform group $G$ must be infinite. The scheme uses two functions: $f\colon G\rightarrow \{0,1\}^*$, which encodes elements of the group as binary strings; and $H\colon\{0,1\}^*\rightarrow G$, a collision-resistant hash function. Using these functions (which are made public along with $G$), a message can be signed and verified as follows:
\begin{itemize}
\item\emph{Key Generation:} The signer first chooses an element $g\in G$, whose centralizer, the set of elements that commute with $g$, contains $1_G$ and powers of $g$ exclusively. The private key consists of $s \in G$ and $n \in \mathbb{N}$, where $n$ is chosen to be highly composite. The public key $x=g^{ns}$ is then published.
\item\emph{Signing Algorithm:} To sign a message $m$, the signer chooses a random element $t\in G$ and a random factorization  $n_in_j$ of $n$, and computes the following (with $||$ denoting concatenation): 
\begin{align*}
y&=g^{n_it}\\
h & = H(m || f(y))\\
\alpha & = t^{-1}shy
\end{align*}
The signature $\sigma=\langle y, \alpha,n_j\rangle$ and the message $m$ are then send to the message recipient.
\item\emph{Verification:} To verify, the recipient computes $h' = H(m || f(y))$, and accepts the message as authentic if and only if the following equality holds:
$$
y^{n_j\alpha} = x^{h'y}.
$$
\end{itemize}

The security of the signature scheme is based on the collision resistance of the hash function, the conjugacy search problem in $G$, and the Diffie-Hellman assumption. Moreover, Alice must maintain a public list of previously used factors of $n$, and regenerate $s$ and $n$ after a few uses.

\subsection{A Key Exchange Using the Subgroup Membership Search Problem}
In \cite{shpilrain2006using}, Shpilrain and Zapata proposed a public key exchange protocol over relatively free groups. Given a free group $G_n$ of rank $n$ and $R \trianglelefteq G_n$,  the quotient group $\mathcal{G}_n=G_n/R$ is \emph{relatively free} if for any endomorphism $\psi$ of $G_n$, $\psi(R)\leq R$.\\

The protocol utilizes two types of automorphisms:
\begin{itemize}
\item Let $\{x_1,\ldots,x_n\}$ be the generators of $\mathcal{G}_n$. The \emph{Neilsen automorphisms} are defined as:
$$
\alpha_j(x_i)=
\left\{
\begin{array}{ll}
x_i^{-1}&i=j\\
x_i&i\neq j
\end{array}
\right.
\beta_{jk}(x_i)=
\left\{
\begin{array}{ll}
x_ix_j&i=k\\
x_i&i\neq k
\end{array}
\right.
$$
\item For relatively free groups like $\mathcal{G}_n$, the Nielsen automorphisms form a subgroup of $\Aut(\mathcal{G}_n)$ under composition. Elements in this subgroup are called \emph{tame} automorphisms. In constructing a private key, the protocol uses both tame and non-tame automorphisms.
\end{itemize}
In the key exchange below, let $\mathcal{F}_n$ and $\mathcal{F}_{n+m}$ denote the relatively free groups of rank $n$ and $n+m$, with respective generating sets $\{x_1,\ldots,x_n\}$ and\\ $\{x_1,\ldots,x_n,x_{n+1},\ldots,x_{n+m}\}$. Moreover, let $\mathcal{F}^i_{j}=\prod_i\mathcal{F}_j$ denote the direct product of $i$ instances of the relatively free group of rank $j$.  Finally, let $z(x_1,\ldots,x_{n+m})$ denote a word $z$ written in the alphabet $\{x_1,\ldots,x_{n+m}\}$. The exchange then proceeds as follows:

\begin{enumerate}
\item Alice chooses an automorphism $\phi \in\Aut(\mathcal{F}_{n+m})$, where $\phi=\tau_1\circ\cdots\circ\tau_k$, a composition of Nielsen automorphisms and non-tame automorphisms which are readily invertible. Alice uses $\phi^{-1}=\tau_k^{-1}\circ\cdots\circ\tau_1^{-1}$ as her private key. For each generator $x_i$ of $\mathcal{F}_{n+m}$, Alice computes the word $\phi(x_i)=y_i(x_1,\ldots, x_{n+m})$. She then computes $\hat{y}_i$, which is the restriction of each $y_i$ to a word in the generators of $\mathcal{F}_{n}$. The tuple $\langle \hat{y}_1,\ldots,\hat{y}_{n+m}\rangle$ is then published as the public key.
\item Bob chooses a word $w$ in the subgroup $S$ of $\mathcal{F}_{n+m}^{n+m}$ consisting of words of the form $v=(v_1(x_1,\ldots,x_n),\ldots,v_n(x_1,\ldots,x_n),1,\ldots,1)$. Thus $S\cong\mathcal{F}_n^n$, and $w=(w_1(x_1,\ldots,x_n),\ldots,w_n(x_1,\ldots,x_n))$. Using the components of the public key, Bob encrypts $w$ by replacing each instance of $x_i$ in $\hat{y}_j$ by $w_i$. The encrypted tuple $\hat{\phi}(w)=\langle \hat{y}_1(w_1,\ldots,w_n),\ldots,\hat{y}_n(w_1,\ldots,w_n)\rangle$ is then sent to Alice.
\item Alice applies $\phi^{-1}$ (restricted to $\mathcal{F}_n^n$) component-wise to $\hat{\phi}(w)$ to recover $w'$, a unique normal form of $w$. This $w'$ is the shared key.
\end{enumerate}

The security of the protocol is two-fold. Decrypting a particular message $\hat{\phi}(w)$ is equivalent to solving the subgroup membership search problem in the subgroup generated by the public key. To recover the private key, an attacker must recover the automorphism $\phi$ and its inverse from the public image of the generators $\hat{y}_i$, restricted to the subgroup $\mathcal{F}_{n}$. Shpilrain and Zapata claim there is no known method of accomplishing this outside of an exhaustive search of $\Aut(\mathcal{F}_{n+m})$.

\

The authors suggest free metabelian groups of rank $r$ (with $r=10, n=8, m=2$) as platform groups for their protocol. Aside from meeting the standard criteria for platform groups, these groups have the requisite supply of non-tame automorphisms and the subgroup membership search problem is known to be super-polynomial in these groups.


\subsection{An Authentication Scheme Based on the Twisted Conjugacy Problem}
\label{twistedauth}
In \cite{shpilrain2008authentication}, Shpilrain and Ushakov introduced a non-commutative authentication scheme based on the Fiat-Shamir scheme. The platform group $G$ can in fact be a semigroup, provided that an antihomomorphism $*:G\rightarrow G$, i.e., $(ab)^{*} = b^{*} a^{*}$, exists. The endomorphism group of $G$ should also be sufficiently large to preclude an exhaustive search. In the simulation of the protocol below, Alice is authenticating herself to Bob:

\begin{enumerate}
\item Alice chooses $s\in G$ as her private key. She then chooses $w,t\in G$ and endomorphisms $\phi,\psi$ such that $t=\psi(s^*)w\phi(s)$. The public key $\langle \phi,\psi,w,t\rangle$ is then published.
\item The commitment/verification exchange proceeds as follows:
\begin{enumerate}
\item Alice chooses an $r\in G$ and computes the \emph{commitment} $u=\psi(r^*)t\phi(r)$, sending it to Bob.
\item Bob chooses a random bit $c$ and sends it to Alice.
\item Alice replies with $v=r$ if $c=0$, and $v=sr$ otherwise.
\item Bob \emph{verifies} the commitment $u$ by computing $u'$, and accepts if $u=u'$:\\
If $c=0$, Bob computes $u'=\psi(v^*)t\phi(v)=\psi(r^*)t\phi(r)$.\\
If $c=1$, Bob computes $u'=\psi(v^*)t\phi(v)$, where
$$
\begin{array}{rl}
u'&=\psi( (sr)^*)t\phi(sr)\\
  &=\psi(r^*)\psi(s^*)w\phi(s)\phi(r)\\
  &=\psi(r^*)t\phi(r).
\end{array}
$$
\end{enumerate}
\end{enumerate}

Note that the commitment/verification steps must be performed $k$ times to yield a probability of successful forgery less than $\frac{1}{2^k}$. The  security of the scheme is based on the apparent hardness of the double twisted conjugacy search problem.

\subsection{Authentication Schemes Based on Semigroup Actions}

Drawing inspiration from the zero-knowledge proof by Feige, Fiat, and Shamir; Grigoriev and Shpilrain \cite{Grigoriev-Shpilrain} introduced two generic protocol schema based upon (semi)group actions and provided several concrete examples.
\subsubsection{An Authentication Scheme Based on the Endomorphism Problem}\label{endoauth}
One such instance of their second protocol is based upon the endomorphism problem. While this scheme can be used with a semigroup or some other algebraic structure, the structure $S$ must meet several criteria:
\begin{itemize}
\item An algorithm exists to determine if function over $S$ is an endomorphism. If $S$ is specified by a presentation this criterion is satisfied by $S$ having an efficiently solvable word problem.
\item An algorithm exists to determine if function over $S$ is an automorphism of $S$.
\item The endomorphism search problem in $S$ should be demonstrably NP-hard.
\end{itemize}

As before, in the protocol exchange below Alice is authenticating herself to Bob:
\begin{enumerate}
\item Alice chooses an endomorphism $\phi\colon S\rightarrow S$ as her private key. Alice then chooses elements $s,t\in S$ such that $t=\phi(s)$. The public key $\langle S,s,t \rangle$ is then published.
\item The commitment/verification exchange proceeds as follows:
\begin{enumerate}
\item Alice chooses an automorphism $\psi$ and computes the \emph{commitment} $u=\psi(t)$, sending it to Bob.
\item Bob chooses a random bit $c$ and sends it to Alice.
\item Alice replies with $v=\psi(t)$ if $c=0$, and $v=\psi\circ\phi$ otherwise.
\item Bob \emph{verifies} the commitment $u$ by computing $u'$:\\
If $c=0$, Bob computes $u'=\psi(t)$ and accepts if  $u=u'$ and $\psi$ is an automorphism.\\
If $c=1$, Bob computes $u'=(\psi\circ\phi)(s)$ and accepts if $u=u'$ and $\psi\circ\phi$ is an endomorphism.
\end{enumerate}
\end{enumerate}

\subsubsection{An Authentication Scheme Based on the Group Isomorphism Problem}\label{isoauth}
The following is a new instance of the first protocol, which requires a class of finitely presented groups $\mathcal{C}$ with the following algorithmic properties:
\begin{itemize}
\item The class $\mathcal{C}$ must have an efficiently solvable isomorphism decision problem.
\item The isomorphism search problem in $\mathcal{C}$ should be demonstrably NP-hard.
\end{itemize}
The protocol exchange is as follows:
\begin{enumerate}
\item Alice chooses two isomorphic groups $G_1$ and $G_2$ from $\mathcal{C}$. Alice then chooses an isomorphism $\alpha\colon G_1\rightarrow G_2$ as her private key, and publishes $\langle G_1,G_2 \rangle$.
\item The commitment/verification exchange proceeds as follows:
\begin{enumerate}
\item Alice chooses a group $G\in\mathcal{C}$ and an isomorphism $\beta\colon G\rightarrow G_1$, sending the \emph{commitment} $G$ to Bob.
\item Bob chooses a random bit $c$ and sends it to Alice.
\item Alice replies with $\gamma=\alpha$ if $c=0$, and $\gamma=\alpha\circ\beta$ otherwise.
\item Bob \emph{verifies} the commitment $G$ by computing $G'=\gamma{G}$:\\
If $c=0$, Bob accepts if $G'\cong G_1$.\\
If $c=1$, Bob accepts if $G'\cong G_2$.

\end{enumerate}
\end{enumerate}

For both of the above authentication schemes, the commitment/verification steps must be performed multiple times to yield a low probability of successful forgery.

\subsection{Secret Sharing Schemes Based on the Word Problem}
Habeeb, Kahrobaei, and Shpilrain \cite{Habeeb-Kahrobaei-Shpilrain} proposed two secret sharing schemes for groups whose presentations satisfy small cancellation conditions. In a $(t,n)$ scheme, the \emph{threshold} $t$ is the number of participants that are required to recover the shared secret (created and disseminated by the ``dealer"), with $n$ the total number of participants.

\

In both schemes, the dealer wishes to share a $k$-bit integer $x$ that will be represented as a column vector $C\in\mathbb{B}^k$.  Prior to initiating the secret sharing, the dealer chooses groups $G_j$ given by the presentations $\langle X | R_j\rangle$, where $X$ is a common generating set and  $R_j$ a unique set of relators for each participant $P_j$. The generating set $X$ is then made public. Note that both schemes require secure communication channels between both the dealer and participants and between the participants themselves. These secure channels can be achieved using any preferred public key exchange protocol.

\subsubsection{An $(n,n)$-threshold Scheme}
In this scheme, all $n$ participants are required to reproduce the secret $x$:
\begin{enumerate}
\item The dealer sends each participant $P_j$ their unique relator set $R_j$.
\item The dealer decomposes $C$ into $n$ vectors $C_j\in\mathbb{B}^k$ such that $C=\sum_jC_j$.
\item Each entry $c_{kj}$ of $C_j$ is then encoded as a word $w_{kj}\in G_j$, such that $w_{kj}\equiv1_{G_j}$ if $c_{kj}=1$ and $w_{kj}\not\equiv1_{G_j}$ otherwise. The $w_{kj}$s are then sent to $P_J$ using an open channel.
\item For each $w_{kj}$, participant $P_j$ solves the word problem in $G_j$ and reconstructs $C_j$.
\item The participants can then recover $C$ by summing over all $C_j$s. Note that a secure sum protocol can be employed so that the $C_j$s need not be divulged to the other participants.
\end{enumerate}
\subsubsection{A $(t,n)$-threshold Scheme}
In this scheme, $t$ participants are required to reproduce the secret $x$. As in Shamir's secret sharing, $x$ must be an element in $\Z_p$ with $p$ prime, and a polynomial $f$ of degree $t-1$ must be chosen by the dealer such that $f(0)=x$. The dealer must also choose $k$-bit integers $y_j\equiv f(j)\;(\mathrm{mod}\;p)$.
\begin{enumerate}
\item The dealer sends each participant $P_j$ their unique relator set $R_j$.
\item Each $y_j$ has its bits $b_{kj}$ encoded as words $w_{kj}\in G_j$ as in the previous scheme.
\item For each $w_{kj}$, participant $P_j$ solves the word problem in $G_j$, yielding $y_j$.
\item The participants can then perform polynomial interpolation using the $y_j$s to recover $f$. The shared secret $x$ is then revealed by evaluating $f(0)$.  If $t\geq 3$, Lagrange interpolation can be employed so that the $B_j$s need not be divulged to the other participants.
\end{enumerate}
The security of these schemes is contingent upon the relators $R_J$ being kept secret.

\section{Cryptanalysis and Attacks}
In this section we present a number of attacks against group-based cryptosystems, with an emphasis on those that are applicable to polycyclic groups.

\subsection{Length-Based Attack}
\label{LBA}
The length-based attack (LBA) is an incomplete, local search that attempts to solve the conjugacy search problem (or its generalized version) by using the length of a word as a heuristic. It was first introduced by Hughes and Tannenbaum \cite{Hughes02length-basedattacks} as a means to attack the AAG key exchange protocol over braid groups. 
In \cite{GarberLBA06}, Garber, Kaplan, Teicher, Tsaban, and Vishne explored the use of length functions based on the  Garside normal form of braid group elements. They demonstrated experimentally that the length-based attack in this context could break the AAG protocol, albeit inefficiently.

\

As the length-based attack is an iterative improvement search, it is susceptible to failing at peaks and plateaux in the search space. In  \cite{MyasnikovUshakov07}, Myasnikov and Ushakov identified when these peaks occur and were able to make successive refinements to the algorithm to yield a high success rate.

\

More recently, the authors of \cite{garber2013analyzing} 
 analyzed the LBA against AAG over polycyclic groups. They found that the success rate of the LBA decreased as the Hirsch length of the platform group increased. Their version of the LBA, essentially a local beam search, is presented below:
\begin{algorithm}[H]
\caption{LBA with Memory 2}
\begin{algorithmic}
\State {Initialize $S=\{(|\overline{b'}|,\overline{b'},1_G)\}$.}
\While{not time out}
	\For {$(|\overline{c}|,\overline{c},x) \in S$}
		\State {Remove $(|\overline{c}|,\overline{c},x)$ from $S$}
		\State {Compute $\overline{c}^{a_i^\varepsilon}$ for all $i \in \{1 \ldots N_1 \}$ and $\varepsilon=\pm 1$}
		\State {\textbf{if} $\overline{c}^{a_i^\varepsilon}=\overline{b}$ \textbf{then} output inverse of $a_i^\varepsilon x$ and stop}
		\State {Save $(|\overline{c}^{a_i^\varepsilon}|,\overline{c}^{a_i^\varepsilon},a_i^\varepsilon x)$ in $S'$}
	\EndFor
	\State {After all conjugation attempts, sort $S'$ by the first element of every tuple}
	\State {Copy the smallest $M$ elements into $S$ and delete the rest of $S'$}
\EndWhile
\State {Otherwise, output FAIL}
\end{algorithmic}
\end{algorithm}
Note that the $a_i$, $b'$, $\bar{b}'$, and $N_1$ are from the AAG protocol exchange in Section \ref{AAG}, while $\bar{c}'$ is a candidate conjugator set. The length of a conjugator set $\bar{c}'=(c_1,\ldots,c_j)$ is defined as $\sum_j|c_j|$.

\subsection{Linear Decomposition Attack}
\label{lineardecomp}
In \cite{myasnikov2015linear}, Miasnikov and Roman'kov introduced the linear decomposition attack. The attack is a general framework for the cryptanalysis of a number of group-theoretic analogues of Diffie-Hellman key exchange. For a protocol to be susceptible to the attack its platform groups must admit a linear representation. Moreover, the algorithmic security assumption of the protocol must be equivalent to commutative linear transformations. Note that the AAG protocol is not susceptible to this attack.

\

Given the linear structure $V$ and subsets $W\leq V$ and $U\leq\End(V)$, the attack first computes a basis for the span of all vectors of the form $w^u$, with $w\in W$ and $u\in\langle U\rangle$. This can be done in polynomial time with respect to the dimension of $V$ and the sizes of $W$ and $U$. This calculation can be performed offline if the platform group for a particular protocol is fixed. The public group elements transmitted during the key exchange can then be decomposed using this basis to reconstruct the shared secret without discovering the private information of each party, negating the need for an attacker to solve the underlying security problem.

\

The attack requires the platform group to be specified by either its linear representation $V$ (as a vector space or an associative algebra) or by a presentation coupled with a faithful embedding into $GL(V)$. Moreover, the linear space into which the group is embedded must be of sufficiently small dimension to make the attack tractable. While the dimension of the smallest linear embeddings of finite groups and some classes of infinite groups such as torsion-free nilpotent and polycyclic-by-finite are known, the authors concede that no such bounds are known for other linear groups, including general polycyclic groups and metabelian groups. 

\subsection{Field-Based Attack}
\label{FBA}
Kotov and Ushakov \cite{KotovUshakov} investigated the security of the AAG key-exchange protocol used with certain polycyclic groups of the form $G_F=\mathcal{O}_F \rtimes U_F$, where $\mathcal{O}_F$ is the maximal order and $U_F$ is the unit group generated by an irreducible polynomial in the algebraic number field $F$. In the semidirect product, $U_F$ acts on $\mathcal{O}_F$ by right multiplication. These groups were the original polycyclic platform groups suggested by Eick and Kahrobaei in \cite{EickKahrobaei04}. In  \cite{garber2013analyzing}, Garber, Kahrobaei, and Lam showed that such groups were resistant to the length-based attack, with the attack's success decreasing as the Hirsch length of the group $G_F$ increased.

\

Contrary to these results, the field-based attack devised by the authors is able to recover the shared key regardless of the group's Hirsch length.  Using a deterministic, polynomial time algorithm, the key is recovered by solving a linear system of conjugacy equations over the field $F$. If the group $G_F$ is specified as a semidirect product and $F$ is given in matrix form, the attack can be directly applied. However, if $G_F$ is given by a polycyclic presentation, the authors construct a linear representation from the presentation prior to recovering the shared key.

\

While the field-based attack is successful in these particular groups, the authors concede that their attack does not preclude other polycyclic groups from consideration for the AAG protocol. We claim that there are other classes of polycyclic groups that are resistant to such an attack. Such platform groups would be specified by their polycyclic presentations and have matrix representations that are not readily computable.

\subsection{Quotient Attack}
In attempting to recover the shared secret from the public information of the AAG protocol, the length-based attack (LBA) operates as if the platform group $G$ is a free group. The success of the LBA on non-free groups motivated Miasnikov and Ushakov in \cite{MyasnikovUshakov} to investigate the asymptotic properties of the given platform groups.  Ultimately they determined that the LBA is successful for groups in which a random choice of elements is very likely to generate a free subgroup of $G$. 

\

These investigations led to a new form of attack for the AAG key exchange protocol and others that use some variation of the membership or conjugacy search problems. Dubbed the quotient attack, the algorithms solve the search problems in a quotient group $G/N$. If $G/N$ possesses the exponentially-generic free basis property the solution in the quotient will yield one in the original group. The time complexity of the attack is contingent upon the particular class of platform groups. For pure braid groups $PB_n$ the authors prove that the complexity is $O(n^2)$.

\

As polycyclic groups do not possess the free basis property nor any free subgroups, this attack is not applicable.

\subsection{Linear Centralizer Attack}
Tsaban \cite{TsabanPoly} devised the linear centralizer attack against AAG over the original braid group platform. The attack exploits a faithful linear representation of a braid group $\mathcal{B}_n$. Using this representation, the algorithm computes a basis for the double centralizer of the public subsets of the AAG protocol (which are contained in their respective double centralizers). This process produces one half of the shared key, after which random elements are tested to find an inverse that yields the other half. The algorithm runs in expected polynomial time with respect to $n$, but is impractical for even modest values of $n$.

\

The applicability of the linear centralizer attack to other platform groups is limited to those whose faithful representations are known and whose linear representations are sufficiently small. As mentioned previously with respect to the linear decomposition attack, these aspects of polycyclic groups are currently unknown.

\section{Conclusion}
\paragraph{}
In this paper we have presented a survey of over ten years of research related to polycyclic group-based cryptography. We began with a study of the algorithmic properties of polycyclic groups. Polycyclic groups admit a number of representations, including polycyclic presentations, multiplication polynomials, and as matrices. In addition to the decidability of the classic decision problems of word, conjugacy, and isomorphism, the twisted conjugacy and  orbit problem are also decidable. Moreover, the conjugacy decision problem for the automorphism group $\Aut(G)$ of a polycyclic group $G$ is decidable.\\
\paragraph{}
We have seen that there are a variety of key exchanges, digital signature systems, and secret sharing schemes for which a polycyclic group is an appropriate choice of platform group. These schemes use several different computational problems in polycyclic groups as listed in the paper, which are beyond use of conjugacy search problem.

\paragraph{}
While there has been considerable research activity concerning polycyclic groups and their attendant cryptosystems over the last decade, many computational complexity and algorithmic questions remain unanswered. We have collected these outstanding problems below, with the hope of stimulating interest in their solutions:
\begin{enumerate}
\item What is the complexity of the isomorphism search problem in polycyclic groups?
\item What is the complexity of the twisted search conjugacy problem in polycyclic groups?
\item What is the complexity of the power conjugacy problem in polycyclic groups?
\item What is the complexity of the geodesic length problem in polycyclic groups?
\item What is the complexity of the $n$-root problem in polycyclic groups?
\item What is the complexity of finding matrix representation of polycyclic groups?
\item What is the complexity of the conjugacy problem in the automorphism of polycyclic groups?
\item What is the complexity of the search endomorphism (automorphism) problem in polycyclic groups?
\item What is the complexity of the homomorphism problem in polycyclic groups?
\item Are polycyclic group-based cryptosystems resistant to quantum algorithms?
\item What is the complexity of the subgroup membership search problem in polycyclic groups?
\end{enumerate}

\section*{Acknowledgements}
We would like to thank Bettina Eick for her contributions regarding polycyclic groups and their algorithmic properties. Delaram Kahrobaei is partially supported by a PSC-CUNY grant from the CUNY Research Foundation, the City Tech Foundation, and ONR (Office of Naval Research) grant N00014-15-1-2164. Delaram Kahrobaei has also partially supported by an NSF travel grant CCF-1564968 to IHP in Paris.
\bibliographystyle{plain}

\bibliography{PCAlgos}
\end{document}